\documentclass[prd,twocolumn,twoside,preprintnumbers,superscriptaddress]{revtex4}

\usepackage{amsmath}
\usepackage{graphicx,axodraw}
\usepackage{dcolumn}
\usepackage[hyperfootnotes=false]{hyperref}
\usepackage{xspace}
\usepackage{color}
\usepackage{url}

\newcommand{\bea}{\begin{eqnarray}}
\newcommand{\eea}{\end{eqnarray}}
\newcommand{\nn}{\nonumber\\}

\newcommand{\eq}[1]{Eq.~\eqref{#1}}

\newcommand{\Br}{\text{Br}}

\begin{document}
%%%%%%%%%%%%%%%%%%%%%%%%%%%%%%%%%
\preprint{\vbox{\hbox{CERN-PH-TH-2015-091}}}
\preprint{\vbox{\hbox{TTP15-018}}}

\title{Lepton-flavour violating \boldmath$B$ decays in generic $Z^\prime$
  models\unboldmath}

\author{Andreas Crivellin}
\affiliation{CERN Theory Division, CH--1211 Geneva 23, Switzerland}

\author{Lars Hofer}
\affiliation{Universitat Aut\`onoma de Barcelona, 08193 Bellaterra,
  Barcelona, Spain}

\author{Joaquim Matias}
\affiliation{Universitat Aut\`onoma de Barcelona, 08193 Bellaterra,
  Barcelona, Spain}

\author{Ulrich Nierste}
\affiliation{Institut f\"ur Theoretische Teilchenphysik, Karlsruhe
  Institute of Technology, 76128
  Karlsruhe, Germany}

\author{Stefan Pokorski} \affiliation{Institute of Theoretical Physics,
  Department of Physics, University of Warsaw}

\author{Janusz Rosiek}
\affiliation{Institute of Theoretical Physics, Department of Physics,
  University of Warsaw}

%%%%%%%%%%%%%%%%%%%%%%%%%%%%%%%%%%
%%%%%%%%%%%%%%%%%%%%%%%%%%%%%%%%%%

\begin{abstract}
  LHCb has reported deviations from the Standard Model in $b\to
  s\mu^+\mu^-$ transitions for which a new neutral gauge boson is a
  prime candidate for an explanation. As this gauge boson has to
  couple in a flavour non-universal way to muons and electrons in
  order to explain $R_K$, it is interesting to examine the possibility
  that also lepton flavour is violated, especially in the light of the
  CMS excess in $h\to\tau^\pm\mu^\mp$. In this article, we investigate
  the perspectives to discover the lepton-flavour violating modes
  $B\to K^{(*)}\tau^\pm\mu^\mp$, $B_s\to \tau^\pm\mu^\mp$ and $B\to
  K^{(*)} \mu^\pm e^\mp$, $B_s\to \mu^\pm e^\mp$. For this purpose we
  consider a simplified model in which new-physics effects originate
  from an additional neutral gauge boson ($Z^\prime$) with generic
  couplings to quarks and leptons. The constraints from $\tau\to3\mu$,
  $\tau\to\mu\nu\bar{\nu}$, $\mu\to e\gamma$, $g_\mu-2$, semi-leptonic
  $b\to s\mu^+\mu^-$ decays, $B\to K^{(*)}\nu\bar{\nu}$ and
  $B_s$--$\overline{B}_s$ mixing are examined. From these decays, we
  determine upper bounds on the decay rates of lepton flavour
  violating $B$ decays. $\rm{Br}(B\to K\nu\bar{\nu})$ limits the branching ratios of
  LFV $B$ decays to be smaller than $8\times 10^{-5} (2\times
    10^{-5})$ for vectorial
  (left-handed) lepton couplings. However, much stronger bounds
  can be obtained by a combined analysis of
    $B_s$--$\overline{B}_s$, $\tau\to3\mu$, $\tau\to\mu\nu\bar{\nu}$ and
    other rare decays. The bounds depend on the amount of  fine-tuning
    among the contributions to $B_s$--$\overline{B}_s$ mixing.  
  Allowing for a fine-tuning at the percent level we find upper
  bounds of the order of $10^{-6}$ for branching ratios into $\tau\mu$
  final states, while $B_s\to \mu^\pm e^\mp$
  is strongly suppressed and only $B\to K^{(*)} \mu^\pm e^\mp$ can be
  experimentally accessible (with a branching ratio of order $10^{-7}$).
\end{abstract}

%%%%%%%%%%%%%%%%%%%%%%%%%%%%%%%%%
%%%%%%%%%%%%%%%%%%%%%%%%%%%%%%%%%
\maketitle

%%%%%%%%%%%
\section{Introduction}
\label{intro}
%%%%%%%%%%%

While most flavour observables agree very well with their Standard-Model (SM) predictions,
there are some exceptions in semi-leptonic $B$ decays (see for example~\cite{Crivellin:2014kga} for a recent review).
LHCb~\cite{Aaij:2014ora} recently found indications for the violation of lepton-flavour universality in the ratio
\begin{equation}
R_K=\dfrac{\Br[B\to K \mu^+\mu^-]}{\Br[B\to K e^+e^-]}=0.745^{+0.090}_{-0.074}\pm 0.036\,,
\end{equation}
which deviates from the theoretically clean SM prediction $R_K^{\rm SM}=1.0003 \pm 0.0001$~\cite{Bobeth:2007dw} by
$2.6\,\sigma$. In addition, LHCb has reported deviations from the SM predictions~\cite{Descotes-Genon:2014uoa,Descotes-Genon:2015xqa,Altmannshofer:2014rta,Straub:2015ica} in the decay $B\to K^*\mu^+\mu^-$ (mainly in an angular observable called $P_5^\prime$~\cite{DescotesGenon:2012zf}) with a
significance of about $3\,\sigma$~\cite{Aaij:2013qta,LHCb:2015dla}. Furthermore, also the measurement of $\Br[B_s\to\phi\mu^+\mu^-]$ disagrees with the SM prediction \cite{Horgan:2013pva, Horgan:2015vla} by about  $3\,\sigma$~\cite{Altmannshofer:2014rta}. 

Interestingly, these discrepancies can be explained in a model-independent approach by a rather large new-physics (NP)
contribution $C_9^{\mu\mu}$ to the Wilson coefficient of the operator $O_9^{\mu\mu}$ (the component of the usual SM operator $O_9$ that
couples to muons, see eq.~(\ref{eq:effHam}))~\cite{Descotes-Genon:2013wba,Altmannshofer:2013foa,Alonso:2014csa,Hiller:2014yaa,Ghosh:2014awa,Altmannshofer:2015sma,inPrep}.  It is encouraging that the value for $C^{\mu\mu}_9$ required to explain $R_K$
(with $C^{ee}_9=0$) is of the same order as the one needed for $B\to K^*\mu^+\mu^-$ and $B_s\to\phi\mu^+\mu^-$
\cite{Hurth:2014vma,Altmannshofer:2014rta}. Taking into account the $3\,$fb$^{-1}$ data for $B\to K^*\mu^+\mu^-$ recently released by the
LHCb collaboration~\cite{LHCb:2015dla}, the global significance is found to be $4.3\,\sigma$ for NP contributing to
$C_9^{\mu\mu}$ only, and $3.13\,\sigma$ in a scenario with $C_9^{\mu\mu}=-C_{10}^{\mu\mu}$ \cite{Altmannshofer:2015sma}.

Many models proposed to explain the $b\to s \mu^+\mu^-$ data contain a heavy neutral gauge boson ($Z^\prime$) which generates a tree-level contribution to $C_9^{\mu\mu}$~\cite{Descotes-Genon:2013wba,Gauld:2013qba,Buras:2013qja,Gauld:2013qja,Buras:2013dea,Altmannshofer:2014cfa}. If the $Z^\prime$ couples
differently to muons and electrons, $R_K$ can be explained simultaneously \cite{Altmannshofer:2014cfa,Crivellin:2015mga,Crivellin:2015lwa, Niehoff:2015bfa,Sierra:2015fma}. Since in this case lepton-flavour universality would be violated, it has been proposed to search for lepton-flavour violating (LFV) $B$ decay modes as well~\cite{Glashow:2014iga}. This is also motivated by the CMS excess in $\Br[h\to\mu\tau]$~\cite{CMS:2014hha} which can be explained simultaneously together with $R_K$, $\Br[B_s\to\phi\mu^+\mu^-]$ and $\Br[B\to K^*\mu^+\mu^-]$ within a single model~\cite{Crivellin:2015mga,Crivellin:2015lwa}.

While the specific model of Refs.~\cite{Crivellin:2015mga,Crivellin:2015lwa} predicts only small effects in LFV $B$ decays, the situation could be different in a generic model. In this article we examine the LFV decays $B\to K^{(*)}\tau^\pm\mu^\mp$, $B_s\to\tau^\pm\mu^\mp$ (and the corresponding $\mu^\pm e^\mp$ channels) studying a simplified model in which the NP effects originate from a heavy new gauge boson $Z^\prime$ of mass $M_{Z^\prime}$ with generic couplings to quarks and leptons \footnote{For an analysis of LFV $B$ decays with leptoquarks see Ref.~\cite{Varzielas:2015iva} and for a model independent
  analysis see Ref.~\cite{Boucenna:2015raa}.}. We introduce the relevant $Z^\prime$ couplings to $\bar s b$ and charged lepton pairs $\ell,\ell^\prime=\tau,\mu,e$ via
\bea
{\cal L}_{Z^\prime} &\supset& \Gamma^L_{\ell\ell^\prime} \bar \ell
\gamma^\mu P_L\ell^\prime +\Gamma^L_{sb} \bar s \gamma^\mu P_L b
+L\leftrightarrow R\,.
\eea
As the $Z'$ is assumed to be much heavier than the scale of
electroweak symmetry breaking, its couplings must respect $SU(2)_L$
gauge invariance. This implies that the couplings to neutrinos and to
left-handed charged leptons are equal:
$\Gamma^L_{\ell_i\ell_j}=\Gamma^L_{\nu_i\nu_j}$\footnote{Here we
  assumed that the $Z'$ boson is a $SU(2)$ singlet and not the neutral
  component of a $SU(2)$ triplet.  In the second case, the relation
  $\Gamma^L_{\ell_i\ell_j}=-\Gamma^L_{\nu_i\nu_j}$ would hold.}.  To
study bounds on the LFV $B$ decay modes, we perform the following
steps:
\newline
1) Motivated by the model-independent fits to $B\to K^* \mu^+\mu^-$,
$B_s\to\phi\mu^+\mu^-$ and $R_K$ we consider two scenarios for the
$Z^\prime$ couplings to leptons: scenario 1 assumes vectorial
couplings, i.e.  $\Gamma^L_{\ell\ell^\prime} =
\Gamma^R_{\ell\ell^\prime} \equiv \Gamma^V_{\ell\ell^\prime}$,
corresponding to $C^{\ell\ell^\prime}_{10} =
C^{\prime\ell\ell^\prime}_{10}=0$.  Scenario 2 considers left-handed
couplings, i.e. $\Gamma^R_{\ell\ell^\prime}=0$, corresponding to
$C_9^{\ell\ell^\prime}= - C_{10}^{\ell\ell^\prime}$. \newline
2) We use the experimental upper bound on $B\to K^{(*)}\nu\bar{\nu}$ decays
to set upper bounds on LFV $B$ decays, independently of the values
of $\Gamma^{L(R)}_{sb}$. \newline
3) From $B_s$--$\overline{B}_s$ mixing we obtain upper limits on
$\Gamma^{L}_{sb}$ as a function of a fine-tuning
measure (to be defined later). \newline
4) In the lepton sector the $Z^\prime$ couplings can be constrained
by $\tau\to 3\mu$ and $\tau\to\mu\nu\bar{\nu}$. \newline
5) Taking into account the constraints 3) and 4) we derive upper limits on the
branching ratios of $B_s\to \tau^\pm\mu^\mp$, $B\to K^{(*)}\tau^\pm\mu^\mp$ which are stronger than the ones obtained
in 2), but depend on the amount of fine-tuning in $B_s$--$\overline{B}_s$ mixing.

\begin{figure}[t]
\begin{center}
\includegraphics[trim = 25 460 145 95,clip,width=0.480\textwidth]{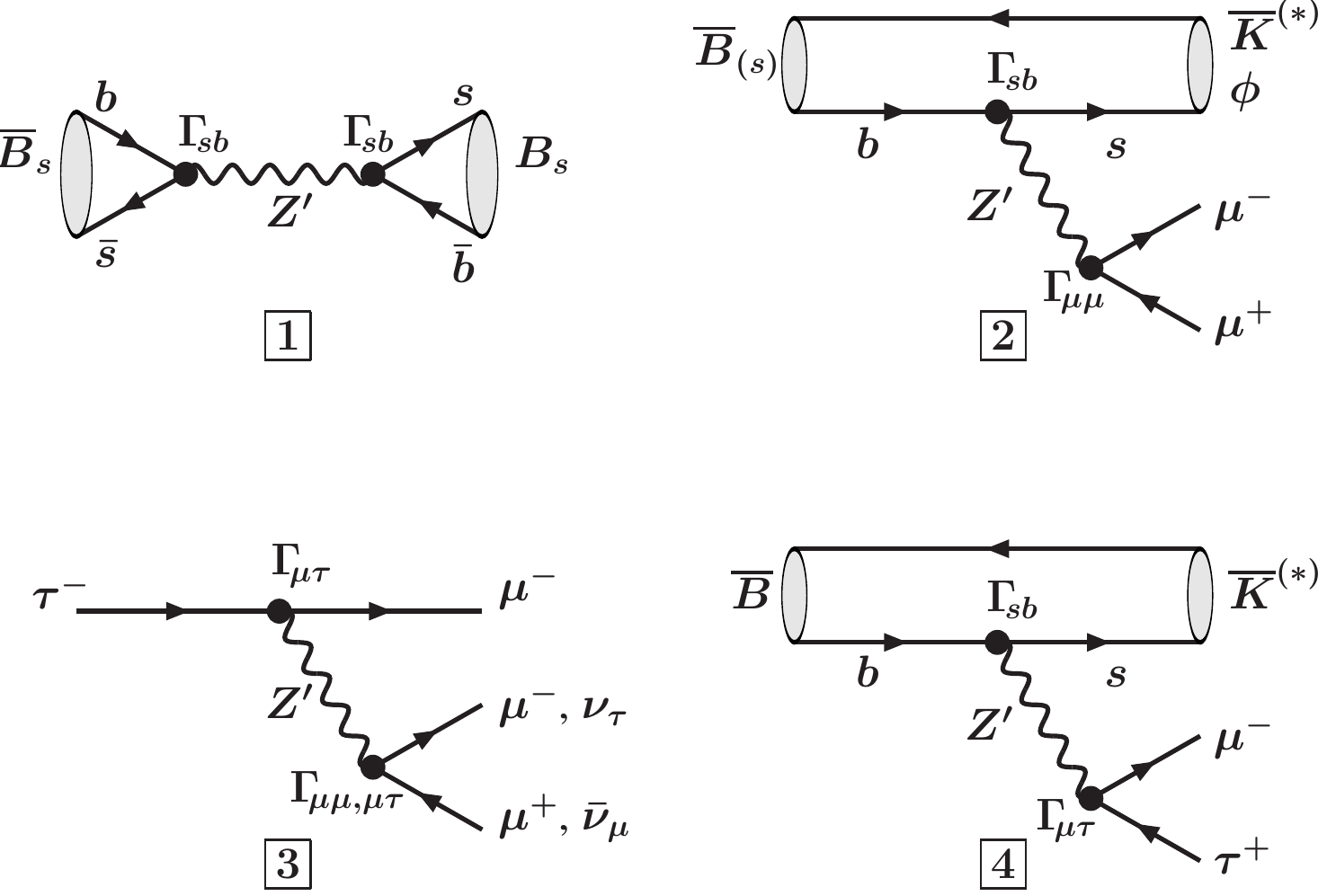}
\end{center}
\caption{Feynman diagrams illustrating the steps 1-4 of our analysis (see text). The diagrams
display the dominant $Z^\prime$ contribution to $\overline{B}_s-B_s$ mixing, 
$\overline{B}\to \overline{K}^{(*)}\mu^+\mu^-$,
$\overline{B}_s\to\phi\mu^+\mu^-$, $\tau\to 3\mu$, $\tau\to\mu\nu\bar{\nu}$ and 
$\overline{B}\to\overline{K}^{(*)}\tau^+\mu^-$.
\label{fig:Feynman}}
\end{figure}
In Fig.~\ref{fig:Feynman} we show the Feynman diagrams for the
dominant $Z^\prime$ contribution corresponding to the steps 1-5 of our
analysis.  We apply a similar procedure to $\mu^\pm e^\mp$ final
states.  In this case the best bounds on the lepton couplings are
coming from $\mu\to e\gamma$ and $\mu\to e\nu\bar{\nu}$.

\section{Processes and observables}
\label{constraints}

In the subsections A-E we collect the formulae for the steps 1-5 of
our analysis outlined in the introduction.

\boldmath
\subsection{$B_s-\overline{B}_s$ mixing\label{susec:A}}
\unboldmath 

Using the notation of Refs.~\cite{Ciuchini:1997bw,Buras:2000if} for
the operators describing $B_s-\overline{B}_s$ mixing, the first
diagram in Fig.~\ref{fig:Feynman} feeds the Wilson coefficients of
\bea
O_1 &=& \left[ \bar s_\alpha \gamma^\mu P_L b_\alpha \right]\left[
  \bar s_\beta \gamma^\mu P_L b_\beta \right] \,, \nn
O_5 &=& \left[ \bar s_\alpha P_L b_\beta \right] \left[ \bar s_\beta
  P_R b_\alpha \right]\,,
\eea
as well as $O_1^\prime$ obtained from $O_1$ by interchanging $P_L\leftrightarrow P_R$.
The coefficients are
\bea
C_1^{(\prime)}={\left(\Gamma^{L(R)}_{sb}\right)^2}/({2 M_{Z^\prime}^2})\,,\;\;\;
C_5={-2\Gamma^{L}_{sb}\Gamma^{R}_{sb}}/({M_{Z^\prime}^2})\,.
\eea
For QCD renormalization group effects we use the next-to-leading order equations
calculated in Refs.~\cite{Ciuchini:1997bw,Buras:2000if}.

\boldmath
\subsection{$b\to s\ell^+\ell^{\prime -}$ transitions
%: \BKs, $B_s\to\mu^+\mu^-$, $B_s\to\phi\mu^{+}\mu^{-}$ and $R_K$
\label{susec:B}}
\unboldmath

For $b\to s\ell^+\ell^{\prime -}$ transitions  we need the operators
\bea
O_{9(10)}^{\ell\ell^\prime} &=& \frac{\alpha }{4\pi}[\bar s{\gamma ^\mu } P_L b]\,[\bar\ell{\gamma _\mu (\gamma^5) }\ell^{\prime}] \,,
\label{eq:effHam}
\eea
and their primed counterparts found by $P_L\leftrightarrow P_R$. $Z^\prime$
contributions to other operators (such as the magnetic operator $O_7$) are
negligible.  The diagrams of Fig.~\ref{fig:Feynman} give
\bea
C_{9,10}^{(\prime)\ell\ell^\prime} &= -\dfrac{\pi}{\sqrt 2  M_{Z^\prime}^2}\dfrac{1}{\alpha G_F V_{tb}
  V_{ts}^\star}\Gamma_{sb}^{L(R)}\left( {\Gamma_{\ell \ell^\prime }^R \pm \Gamma_{\ell \ell^\prime }^L} \right)\,,
%
%C_{10}^{(\prime)\ell\ell^\prime} &= -\dfrac{\pi}{\sqrt 2  M_{Z^\prime}^2}\dfrac{1}{\alpha G_F V_{tb}  V_{ts}^\star}\Gamma_{sb}^{L(R)}\left( {\Gamma_{\ell \ell^\prime }^R-
\label{eq:c9}
\eea
which have to be multiplied by $-4G_F V_{tb}V_{ts}^*/\sqrt{2} $ in the effective Hamiltonian.

As first noted in Ref.~\cite{Descotes-Genon:2013wba,
  Descotes-Genon:2013zva} a good fit to $B\to K^*\mu^+\mu^-$ data,
{leaving ${\rm Br}[B_s\to\mu^+\mu^-]$ unchanged,} is obtained with
$C^{\mu\mu}_9<0$ and $C^{\prime\mu\mu}_9,C^{(\prime)\mu\mu}_{10}\sim
0$.  Another interesting solution is given by $C_{9}^{\mu\mu} =
-C_{10}^{\mu\mu}$~\cite{Altmannshofer:2014rta,Altmannshofer:2015sma}.
  
In our analysis we use the global fit of Ref.~\cite{Altmannshofer:2014rta,Altmannshofer:2015sma}, resulting for the two scenarios under consideration in
\bea
-0.53\, (-0.81) \geq & C_9^{\mu\mu}& \geq(-1.32)\, -1.54
\,, \label{scenario1}\;\;\\
-0.18\, (-0.35) \geq & C_{9}^{\mu\mu}=-C_{10}^{\mu\mu}& \geq(-0.71)\;
-0.91 \,,\;\;\label{scenario2}
\label{eq:C9fit}
\eea
at the ($1\,\sigma$) $2\,\sigma$ level, respectively. The quoted ranges are in
good agreement with preliminary results of Ref.~\cite{inPrep}. Note that${\rm Br}[B_s\to\mu^+\mu^-]$ is suppressed in scenario 2 compared to the SM. This effect is taken into account via the global fit used in our analysis.   

\boldmath
\subsection{$B\to K^{(*)}\nu\bar{\nu}$}
\unboldmath

Following~\cite{Buras:2014fpa} we write the relevant effective Hamiltonian as
\begin{align}
H_{\rm eff}^{\nu\nu^\prime} &= - \frac{4 G_F }{\sqrt 2 } V_{tb},
V_{ts}^*\left( {{C_L^{\nu\nu^\prime}}{O_L^{\nu\nu^\prime}} + {C_R^{\nu\nu^\prime}}{O_R^{\nu\nu^\prime}}}
\right)\,\\ O_{L,R}^{\nu\nu^\prime} &= \frac{\alpha }{{4\pi }} [\bar s{\gamma
    ^\mu }{P_{L,R}}b][{{\bar \nu }}{\gamma _\mu }\left( {1 - {\gamma
      ^5}} \right){\nu^\prime}]\,,\\
C_{L(R)}^{\nu\nu^\prime} &= -\dfrac{\pi}{\sqrt 2  M_{Z^\prime}^2}\dfrac{1}{\alpha G_F V_{tb}
  V_{ts}^\star}\Gamma_{sb}^{L(R)}{\Gamma_{\nu \nu^\prime }^L}\,.
\end{align}
In the approximation % of vanishing right-handed
$\Gamma_{sb}^{R}=0$, the branching ratio (normalized to the SM prediction) reads
\begin{equation}
R_{{K^{(*)}}}^{\nu \bar \nu } = \frac{1}{3}\sum\limits_{i,j = 1}^3 {{{{{\left| {C_L^{ij}} \right|}^2}} \mathord{\left/
 {\vphantom {{{{\left| {C_L^{ij}} \right|}^2}} {{{\left| {C_L^{{\rm{SM}}}} \right|}^2}}}} \right.
 \kern-\nulldelimiterspace} {{{\left| {C_L^{{\rm{SM}}}} \right|}^2}}}}\,,\label{RKnunu}
\end{equation}
with $C_L^{\rm SM}\approx-1.47/s_W^2\approx-6.4$. The complete expressions can be found in Ref.~\cite{Buras:2014fpa}. The current experimental limits are ${R_K^{\nu\bar{\nu}}} < 4.3$~\cite{Lees:2013kla} and ${R_{{K^*}}^{\nu\bar{\nu}}} < 4.4$~\cite{Lutz:2013ftz}. 

Due to $SU(2)$ invariance, we have $C_{L}^{ij} =
(C_{9}^{ij}-C_{10}^{ij})/2$, so that $C_{L}^{ij}= C_{9}^{ij}/2$ in
scenario~1 and $C_{L}^{ij}=C_{9}^{ij}$ in scenario~2.  

\boldmath
\subsection{$\tau\to\mu\nu\bar{\nu}$, $\mu\to e\nu\bar{\nu}$ and $\tau\to 3\mu$}
\label{susec:C}
\unboldmath The $Z^\prime$ boson contributes to $\tau\to\mu\nu\bar{\nu}$ in two
ways: it generates loop corrections to the $W$ exchange diagram (as in the
lepton-flavour conserving case \cite{Altmannshofer:2014cfa}) and it mediates
$\tau\to\mu\nu\bar{\nu}$ at tree-level via LFV couplings.  The latter
contribution decouples as $1/m_{Z'}^2$ from the branching ratio ${\rm Br}\left[ \tau \to \mu \nu \bar{\nu}\right]$ for $\nu_\tau\bar{\nu}_\mu$ final-states where it interferes with the SM tree-level amplitude, and as $1/m_{Z^\prime}^4$ for other final-state flavours $\nu_i\bar{\nu}_j$. We find
\begin{widetext}
 \begin{equation}
{\rm Br}\left[ {\tau \to \mu \nu \bar\nu } \right] \!=\! 
{\rm Br}\left[ {\tau   \to \mu \nu \bar\nu } \right]_{SM}
\left(1+\dfrac{3  \Gamma^L_{\mu\mu}\Gamma^L_{\tau\tau}}{4\pi^2} \frac{\log
    {m_W^2}/{m_{Z'}^2}}{1-{m_{Z'}^2}/{m_W^2}}\right) 
-\, \frac{ 8G_F m_\tau^5 }{ 1536 \sqrt{2} \pi ^3\Gamma_\tau m_{Z'}^2 }
   \mathop{\rm Re} \nolimits \left[ {\Gamma _{\mu \tau }^L
        \Gamma _{\nu _\tau \nu _\mu }^L } \right] 
 + {\cal O} \left( \frac{1}{m_{Z'}^4} \right). \,  
\label{eq:tauvv}
\end{equation} %%%
\end{widetext}
The HFAG value \cite{Amhis:2014hma} for the branching ratio reads%
\bea
\label{eq:PDGtau}
\text{BR}(\tau \to \mu \nu_\tau \bar \nu_\mu)_\text{exp} = (17.39 \pm 0.04)\% \,.
\eea
This should be compared to
\bea
\text{BR}(\tau \to \mu \nu_\tau \bar \nu_\mu)_\text{SM} = (17.29 \pm 0.03) \% \,,
\eea
obtained from the SM prediction in Ref.~\cite{Pich:2013lsa} and a combination of the $\tau$ lifetime measurements in
Refs.~\cite{Belous:2013dba,Alexander:1996nc, Barate:1997be, Acciarri:2000vq,Abdallah:2003yq,Balest:1996cr}. The difference is given by
\bea
\Delta_{\tau\to\mu\nu\bar{\nu}} &\equiv& \Br(\tau \to \mu \nu_\tau \bar
\nu_\mu)_\text{SM}-\Br(\tau \to \mu \nu_\tau \bar \nu_\mu)_\text{exp}
\nn
&=& (-1.0 \pm 1.1) \times 10^{-3}\,.
\eea
at the $2\,\sigma$ level, adding the error originating from the SM theory predictions linear to the experimental one. In the analogous case of $\Gamma_{\mu e}$ we demand
\bea
|\Delta_{\mu\to e\nu\bar{\nu}}| \le 4 \times 10^{-5}\,.
\eea
This choice restricts corrections to the Fermi-constant, defined through the
decay $\mu\to e\nu\bar{\nu}$, to the sub per-mille level and thereby avoids
conflicts with electroweak precision data.

The $Z^\prime$ boson further mediates the LFV three body decay
$\tau \to 3\mu$ at tree-level, with the branching ratio given by (cf.\
e.g.~\cite{Langacker:2008yv,Crivellin:2013hpa})
\bea
\Br\left[ \tau \to 3\mu \right] = \dfrac{m_\tau^5}{1536 \pi^3
  \Gamma_\tau M_{Z^\prime}^4} \left[ 2\left( \left| \Gamma _{\mu
	    \tau }^L \Gamma _{\mu \mu }^L \right|^2 \right.\right. \nn 
+ \left.\left.  \left|
	 \Gamma _{\mu \tau }^R\Gamma _{\mu \mu }^R \right|^2 
  \right) + \left| \Gamma _{\mu \tau }^L\Gamma _{\mu \mu }^R
      \right|^2 + \left| \Gamma _{\mu \tau }^R\Gamma _{\mu \mu
	}^L \right|^2 \right]\,.
\label{eq:tau3mu}
\eea
Combining Belle ~\cite{Aubert:2009ag} and BaBar~\cite{Lees:2010ez} data gives $
\Br\left[ \tau \to 3\mu \right] \leq 1.2\times 10^{-8}$ at
$90\%$~C.L.~\cite{Amhis:2014hma}.  The corresponding decay $\mu\to 3e$ does not
affect our phenomenology, because it involves $\Gamma_{ee}$ which we set to zero
to comply with $R_K$.

\subsection{Lepton-flavour violating \boldmath$B$\unboldmath\ decays}
\label{susec:H}
Here we give formulas for the branching ratios of LFV $B$ decays, taking into
account the contributions from the operators $O_{9}^{(\prime)\ell\ell^\prime}$
and $O_{10}^{(\prime)\ell\ell^\prime}$ relevant for our model. For
$B_s\to\ell^+\ell^{\prime -}$ (with $\ell\neq \ell^\prime$) we use the results
of Ref.~\cite{Dedes:2008iw} neglecting the mass of the lighter lepton. The
branching ratios for $B\to K^{(*)}\tau^\pm\mu^\mp, B\to K^{(*)} \mu^\pm e^\mp$
are computed using form factors from Ref.~\cite{Bouchard:2013eph} (see also
Refs.~\cite{Horgan:2015vla,Horgan:2013hoa}). The results read
\begin{widetext}
  \bea \Br\left[ B_s \to \ell^+ \ell^{\prime-} \right] &=& \dfrac{\tau_{B_s}
    {\rm
        Max}[m_\ell^2,m_{\ell^\prime}^2] M_{B_s} f_{B_s}^2}{64\pi^3} \alpha^2 G_F^2 \left| V_{tb}
    V_{ts}^*\right|^2 \left(1 - \dfrac{{\rm
        Max}[m_\ell^2,m_{\ell^\prime}^2]}{M_{B_s}^2}\right)^2\nonumber\\
				&&\times\left(\left|
      C_{9}^{\ell \ell^\prime}-C^{\prime\ell \ell^\prime}_{9}\right|^2 + \left|
      C_{10}^{\ell \ell^\prime}-C^{\prime\ell \ell^\prime}_{10} \right|^2
  \right) \,,\,\quad\,
\label{bstaumu}\\
\Br[B\to K^{(*)}\ell^+\ell^{\prime-}] &=& 10^{-9} \left(a_{K^{(*)}\ell\ell^\prime}
\left|C_9^{\ell\ell^\prime} + C_9^{\prime\ell\ell^\prime} \right|^2 +
b_{K^{(*)}\ell\ell^\prime}\left|C_{10}^{\ell\ell^\prime} +
C_{10}^{\prime\ell\ell^\prime} \right|^2 \right. \nn && +\left.
c_{K^{(*)}\ell\ell^\prime}\left|C_9^{\ell\ell^\prime}
-C_9^{\prime\ell\ell^\prime} \right|^2 +
d_{K^{(*)}\ell\ell^\prime}\left|C_{10}^{\ell\ell^\prime}
-C_{10}^{\prime\ell\ell^\prime} \right|^2 \right)\,,
\label{bkstaumu}
\eea
with
\begin{center}
\begin{tabular}{|c|c|c|c|c|c|c|c|c|}
\hline
$\ell\ell^\prime $ & $a_{K\ell\ell^\prime}$ & $b_{K\ell\ell^\prime}$ &
$c_{K\ell\ell^\prime}$ & $d_{K\ell\ell^\prime}$ &
$a_{K^*\ell\ell^\prime}$ & $b_{K^*\ell\ell^\prime}$ &
$c_{K^*\ell\ell^\prime}$ & $d_{K^*\ell\ell^\prime}$ \\
\hline
$\;\tau\mu\;$ & $\;9.6 \pm 1.0\;$ & $\;10.0 \pm 1.3\;$ & $0$ & $0$ & $\;3.0 \pm
0.8\;$ & $\;2.7 \pm 0.7\;$ & $\;16.4 \pm 2.1\;$ & $\;15.4 \pm 1.9\;$
\\
$\mu e$ & $15.4 \pm 3.1$ & $15.7 \pm 3.1$ & $0$ & $0$ & $5.6 \pm 1.9$ & $5.6 \pm
1.9$ & $29.1 \pm 4.9$ & $29.1 \pm 4.9$ \\
\hline
\end{tabular} .
\end{center}
\end{widetext}

Note that the results\footnote{Our predictions are for $B^0\to K^{(*)0}\ell^+
\ell^{\prime-}$, those for the charged modes $B^+\to
K^{(*)+}\ell^+ \ell^{\prime-}$ can be found by multiplying by the ratio 
$\tau_{B^+}/\tau_{B^0} $ of  $B$-meson lifetimes.} in Eqs.~(\ref{bstaumu}) and (\ref{bkstaumu}) are for 
$\ell^-\ell^{\prime+}$ final states and not for the sums
$\ell^\pm\ell^{\prime\mp}=\ell^-\ell^{\prime+}+\ell^+\ell^{\prime-}$ constrained 
experimentally~\cite{Amhis:2014hma}:
\bea
{\rm Br}\left[B^+\to K^+\tau^\pm\mu^\mp \right]_{\rm exp} &\le& 4.8\times 10^{-5}	 \,,\nonumber\\
{\rm Br}\left[B^+\to K^+\mu^\pm e^\mp \right]_{\rm exp} &\le& 9.1\times 10^{-8}   \,,\nonumber\\
{\rm Br}\left[B\to K^*\mu^\pm e^\mp\right]_{\rm exp} &\le& 1.4\times 10^{-6} \,,\nonumber\\
{\rm Br}\left[B_s\to \mu^\pm e^\mp \right]_{\rm exp} &\le& 1.2\times 10^{-8}   \,. \label{BllEXP}
\eea

\section{Phenomenological analysis}

\begin{figure}[t]
\begin{center}
\begin{tabular}{cp{7mm}c}
\includegraphics[width=0.400\textwidth]{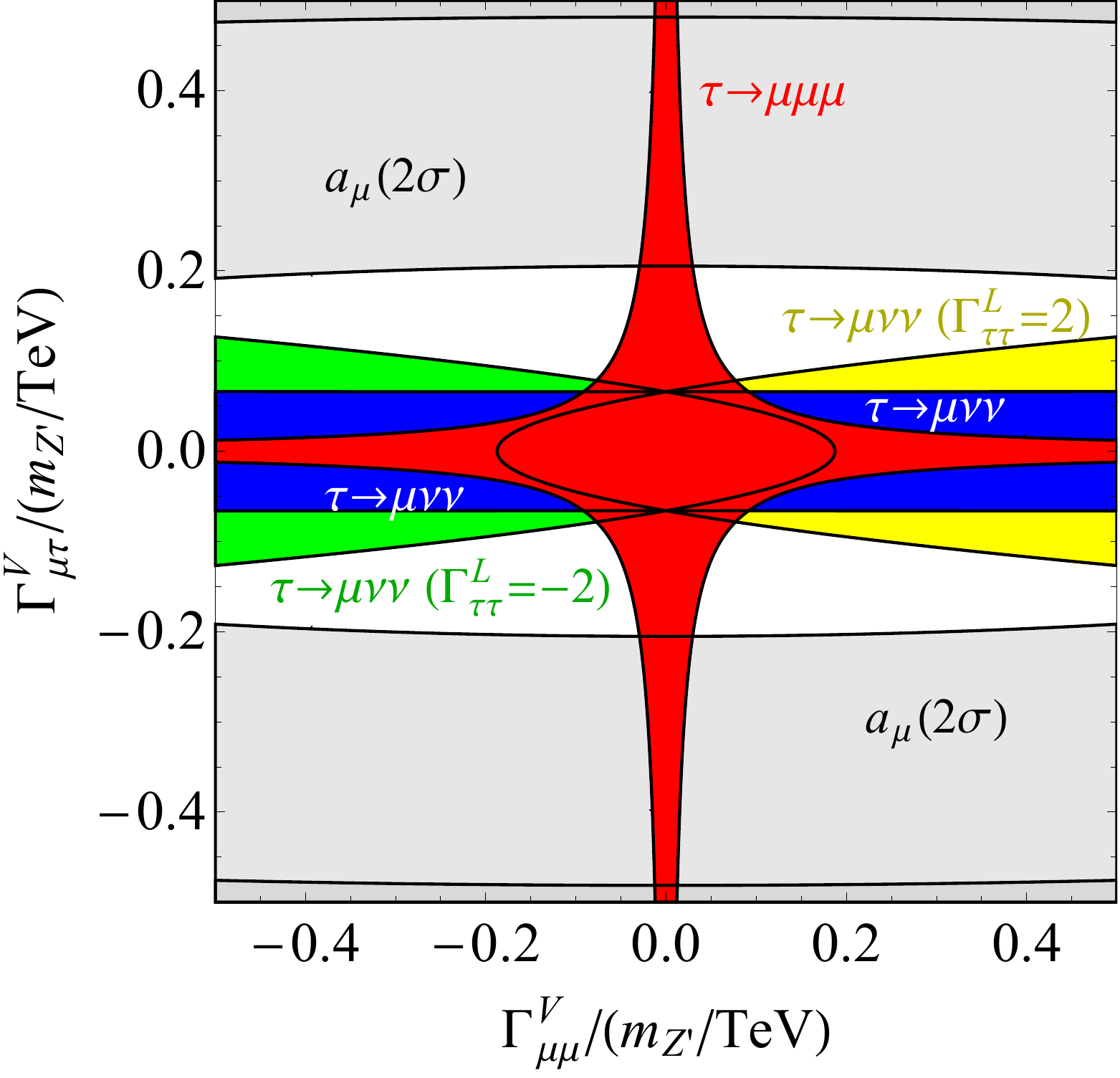}
\end{tabular}
\end{center}
\caption{Allowed $2\,\sigma$ regions in the $\Gamma^{V}_{\mu\mu}-\Gamma^{V}_{\mu\tau}$ plane from $\tau\to\mu\nu\bar{\nu}$ for $\Gamma^{V}_{\tau\tau}=0$ (blue), $\Gamma^{V}_{\tau\tau}=-2$ (yellow), $\Gamma^{V}_{\tau\tau}=2$ (green), $\tau\to3\mu$ (red) and $a_{\mu}$ (light grey) for $m_{Z^\prime}=1\,{\rm TeV}$. The dependence of the bounds on the Z' mass is only logarithmic.  Although NP effects move $a_\mu$ to the right direction, it cannot be explained within our model and we do not impose it as a constraint later on in our analysis.
\label{LeptonConstraints}}
\end{figure}

First of all, one can already derive an upper limit on LFV $B$
  decays from $B\to K\nu\bar{\nu}$ alone, simply by employing gauge
  invariance~\footnote{As stated before, we assume that the $Z'$ is a
    $SU(2)_L$ singlet. The same upper bound from $B\to K\nu\bar{\nu}$ would
    also apply if the $Z'$ would be the neutral component of a $SU(2)_L$
    triplet, but would not hold anymore if it is a mixture of different
    representations.}. As one can see from \eq{RKnunu} the contribution
  for LFV couplings can only be positive. Therefore we can give a strict
  upper limit on $|C_9^{\mu\tau}|$ assuming that all other contributions
  vanish~\footnote{This limit would be even slightly stronger if one
    would assume a vanishing NP contribution in the $ee$ sector and a small
    contribution to $\mu\mu$ (as preferred by the global fit) together
    with a maximally destructive interference in $\tau\tau$.}. We obtain
  $|C_9^{\mu\tau}|\leq 46$ for our scenario 1 and
  $|C_9^{\mu\tau}|=|C_{10}^{\mu\tau}|\leq 23$ for scenario 2. This
  results in upper limits on the branching ratios of $b\to s\tau\mu$
  decays:
%\begin{widetext}
\begin{align}
% {\rm Scenario~1\,(2):\;\;}
{\rm Br}[B\to K^*\tau\mu]& \approx {\rm Br}[B_s\to \tau\mu] \approx 2 
  {\rm Br}[B\to K\tau\mu]\nonumber \\ 
   & < 
    \left\{ \begin{array}{ll}8\times 10^{-5} & \mbox{~~in scenario 1,}\\
                            2\times 10^{-5} & \mbox{~~in scenario 2}.
           \end{array} \right.
\end{align}
%\end{widetext}
However, as we will show now, even stronger constraints can be obtained by employing the combined constraints from the other observables. Let us first examine the numerical impact of the leptonic constraints.
As seen from Fig.~\ref{LeptonConstraints}, for our scenario~1 (vectorial couplings), $\tau\to\mu\nu\bar{\nu}$ 
rules out an explanation of $a_{\mu}$ via a non-vanishing $\Gamma^V_{\mu\tau}$ (contrary to claims in
Ref.~\cite{Huang:2001zx} where $\tau\to \mu\nu\bar{\nu}$ was not considered).
The constraints from $Z\to\mu^+\mu^-$ and $Z\to\tau^\pm\mu^\mp$ as well as from neutrino-trident production (NTP) 
(see Ref.~\cite{Altmannshofer:2014pba}) are irrelevant in the displayed $\Gamma_{\mu\mu}$--$\Gamma_{\mu\tau}$ region for the
considered $Z^\prime$ masses (around 1 TeV and above). The situation is similar in scenario~2 
(left-handed couplings). In this case the interference with the SM terms in $a_{\mu}$ is always
destructive, albeit small. 

The most stringent constraints on the couplings $\Gamma^{L,R}_{bs}$ stem from
$B_s-\overline{B}_s$ mixing. Using the 95$\%$ CL results on $\Delta m_{B_s}$ by
the UTfit collaboration \cite{UTfit,Bona:2006sa,Bona:2007vi}\footnote{Similar
 results are obtained by the CKMfitter collaboration \cite{Charles:2004jd}.}
one obtains
\bea
- 0.10<   \Delta R_{B_s} \equiv {\Delta m_{B_s}}/{\Delta m_{B_s}^{SM}} -1 <0.23\, .\label{eq:gblim}
\eea
One can now derive limits on $\Gamma^{L}_{sb}$ and $\Gamma^{R}_{sb}$ via the
relation
\bea
\Delta R_{B_s} = \dfrac{a_{B_s}}{M_{Z^\prime}^2} \left[(\Gamma^{L}_{sb})^2 + (\Gamma^{R}_{sb})^2 - b_{B_s}
  \Gamma^{L}_{sb} \Gamma^{R}_{sb} \right] \,.
\label{eq:rdef}
\eea
The coefficients $a_{B_s}, b_{B_s}$ only exhibit a weak logarithmic dependence
on $M_{Z^\prime}$ (about 3\% when varying $M_{Z^\prime}$ from 1 to 3 TeV) and we
 use the values at $M_{Z^\prime}=1$ TeV:
\bea 
a_{B_s}/{M_{Z^\prime}^2} \approx 5700~\mathrm{TeV^{-2}}\,,\qquad b_{B_s} \approx 8.8\,.  
\eea
\begin{figure}[t]
\includegraphics[width=0.400\textwidth]{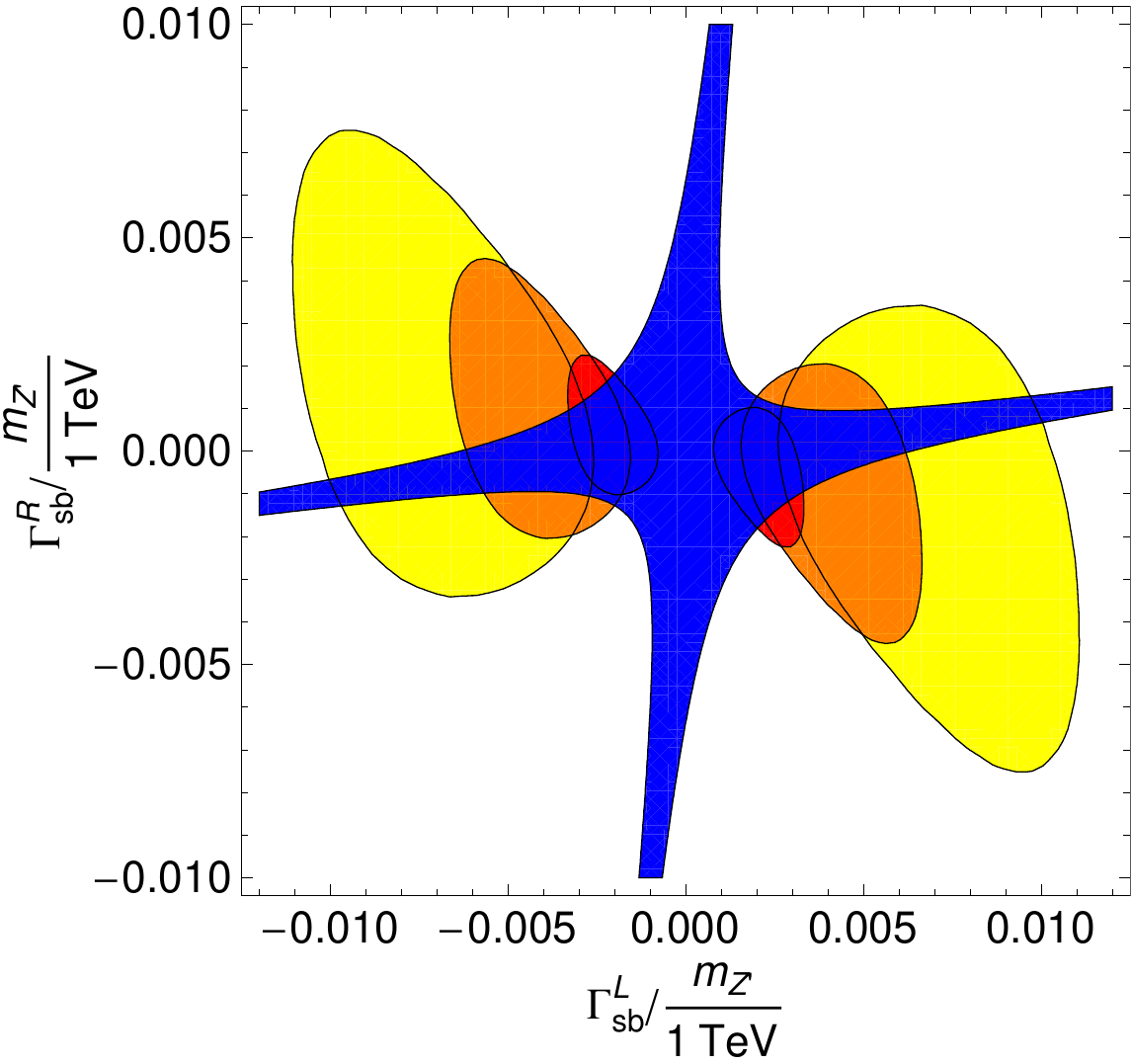}
\caption{Allowed regions in the 
  $\Gamma^{L}_{sb}/M_{Z^\prime}- \Gamma^{R}_{sb}/M_{Z^\prime}$ plane from $B_s$-$\overline{B}_s$ mixing
  (blue), and from the $C^{\mu\mu}_9-C^{(\prime)\mu\mu}_9$ fit of  Ref.~\cite{Altmannshofer:2014rta} to $B\to K^*\mu^+\mu^-$, 
  $B_s\to\phi\mu^+\mu^-$ and $R_K$, with $\Gamma^V_{\mu\mu}=\pm 1$  (red), $\Gamma_{\mu\mu}^V=\pm 0.5$ (orange) and  $\Gamma^V_{\mu\mu}=\pm0.3$ (yellow).	Note that the allowed regions  with positive (negative) $ \Gamma^{L}_{sb}$ correspond to positive  (negative) $\Gamma^V_{\mu\mu}$. The bounds are shown for $m_{Z'}=1$ TeV but their dependence on the Z'  mass is only logarithmic.
\label{Bs-mixingC9}}
\end{figure}
The bounds resulting from Eqs.~(\ref{eq:gblim}) and (\ref{eq:rdef}) (shown by the
blue contour of Fig.~\ref{Bs-mixingC9}) are weakened if $\Gamma^{L}_{sb}$ and $\Gamma^{R}_{sb}$ have the same sign with
$|\Gamma^{R}_{sb}|\ll|\Gamma^{L}_{sb}|$ or $|\Gamma^{R}_{sb}|\gg|\Gamma^{L}_{sb}|$, as a consequence of cancellations in
eq.~(\ref{eq:rdef}). At the $2\,\sigma$ level, current $b\to s\mu^+\mu^-$ data requires a substantial non-zero contribution to $C_9^{\mu\mu}$, eliminating the
option $|\Gamma^{R}_{sb}|\gg|\Gamma^{L}_{sb}|$.  Fig.~\ref{Bs-mixingC9} illustrates the combined constraints from $b\to s\mu^+\mu^-$ data~\cite{Altmannshofer:2014rta,Altmannshofer:2015sma} for different values of $\Gamma^V_{\mu\mu}$ (scenario 1). In principle there is no
upper limit on $|\Gamma^{L}_{sb}|$ as long as $b\to s\mu^-\mu^-$ data permits small but non-vanishing contributions to the primed operators
$C_9^\prime$ and/or $C_{10}^\prime$\footnote{Likewise there is no upper limit on $\Gamma_{sb}$ if the $Z'$ does not couple to muons, as constraints from $b\to
  s \mu^+ \mu^-$ transitions do not apply in this case.}. Therefore we quantify the degree of cancellation in Eq.~(\ref{eq:rdef}) by the
following fine-tuning measure:
\bea
X_{B_s} &=&\dfrac{(\Gamma_{sb}^L)^2 + (\Gamma_{sb}^R)^2 + b_{B_s}  \Gamma_{sb}^L \Gamma_{sb}^R}{ (\Gamma_{sb}^L)^2 +(\Gamma_{sb}^R)^2 -
  b_{B_s} \Gamma_{sb}^L \Gamma_{sb}^R}\nn
& =& \dfrac{2a_{B_s}}{M_{Z^\prime}^2\Delta  R_{B_s}}\left[(\Gamma_{sb}^L)^2 + (\Gamma_{sb}^R)^2\right] - 1 \,,
\label{eq:ft}
\eea
Restricting $X_{B_s}$ to an acceptable value limits the maximal size
$|\Gamma_{sb}^L|$.  As we are exclusively interested in scenarios with
$C_{9,10}^{\mu\mu}\gg C_{9,10}^{\prime\mu\mu}$, we neglect
$(\Gamma_{sb}^R)^2$ in \eq{eq:ft} and express $\Gamma_{sb}^L$ in terms of
$X_{B_s}$ and $\Delta R_{B_s}$ as
\bea {\left|\Gamma_{sb}^L\right|}/{M_{Z^\prime}}\!=\!\sqrt{{\Delta
    R_{B_s} \left(1 + X_{B_s} \right) }/{(2a_{B_s})}} \!\leq\! c_{B_s} \sqrt {1 +
  X_{B_s } }\,.  \nonumber
\eea

Note that we take all couplings $\Gamma^{L,R}_{\ell\ell^\prime}$ real to comply
with CP data in $B_s-\overline{B}_s$ mixing. Using the maximal $|\Delta
R_{B_s}|$ allowed by Eq.~(\ref{eq:gblim}), we find
\bea c_{B_s}={\rm max\left[\sqrt{\Delta R_{B_s}/2a_{B_s}}\right]} \approx
0.0045~\mathrm{TeV}^{-1} \,.  \eea
\begin{figure}[thb]
\begin{center}
\begin{tabular}{cp{7mm}c}
\includegraphics[width=0.400\textwidth]{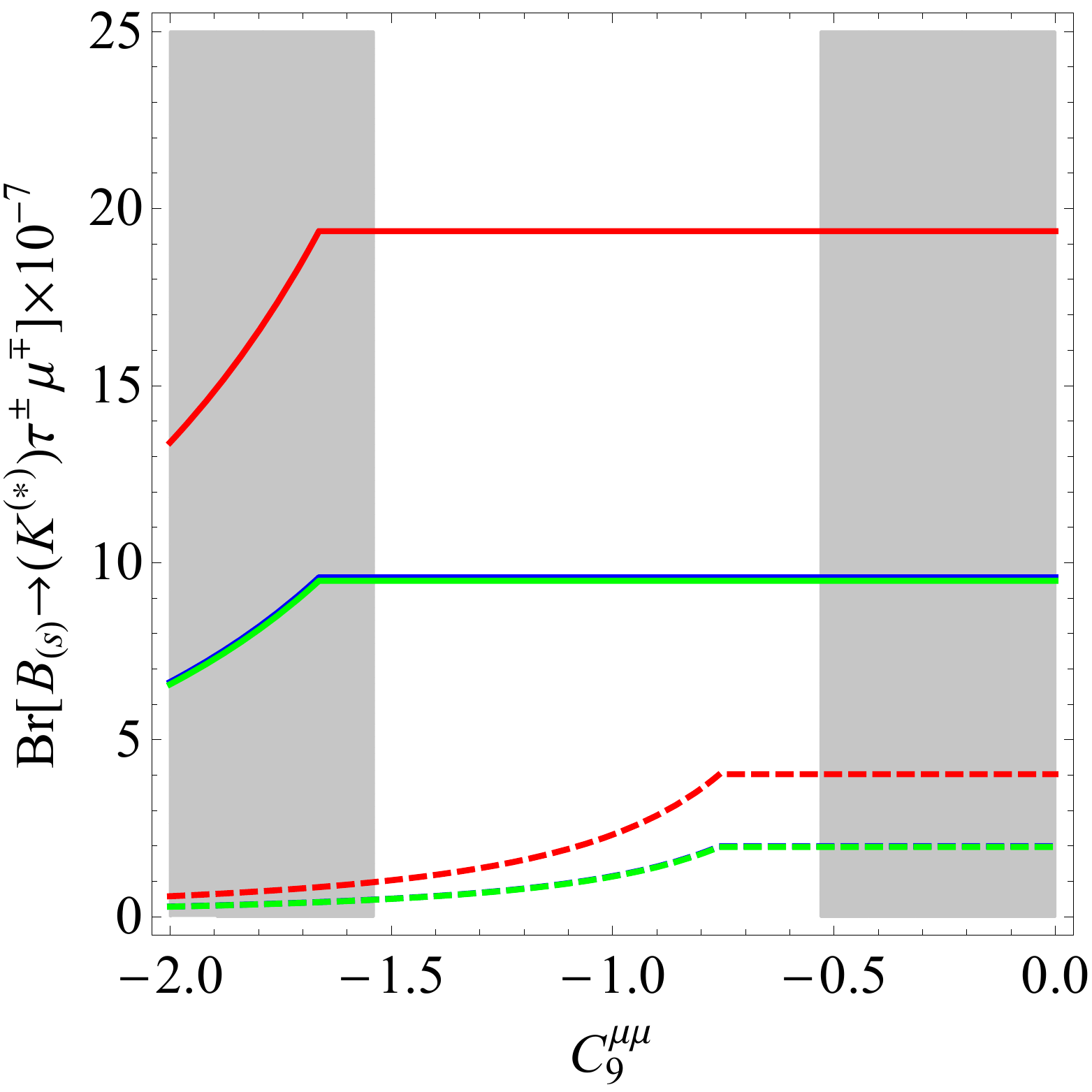}
\end{tabular}
\end{center}
\caption{Maximal value of ${\rm Br}[B\to K^*\tau^\pm\mu^\mp]$ (red),
  ${\rm Br}[B\to K\tau^\pm\mu^\mp]$ (blue) and ${\rm Br}[B_s\to \tau^\pm\mu^\mp]$
  (green) in scenario 1 as a function of $C_9^{\mu\mu}$ for a
  fine-tuning of $X_{B_s}=100$ (solid lines) and $X_{B_s}=20$ (dashed lines).  The bounds are shown for $m_{Z'}=1$ TeV but their dependence on the Z'  mass is only logarithmic.\label{Brmax} }
\end{figure}
Combining the bound on $\Gamma_{bs}^L$ and Eqs.~(\ref{eq:tauvv}),(\ref{eq:tau3mu}) we derive upper limits for the
coefficient $C_9^{\mu\tau}$:
\bea
{\left| {C_9^{\mu \tau }} \right|^2} &\le& A_{3\mu}\,
\dfrac{{64 {\pi ^7}{\Gamma _\tau }{c_{{B_s}}^4}}}{{m_\tau ^5{\alpha
      ^4}G_F^4{{\left| {{V_{tb}}V_{ts}^ \star } \right|}^4}}}\times \nn
&&
{\rm max}\{{\rm{Br}}{\left[ {\tau \to 3\mu } \right]_{\exp
}}\}\times
\dfrac{{{{\left( {1 + {X_{{B_s}}}}
	\right)}^2}}}{{{{\left| {C_9^{\mu \mu }}
	\right|}^2}}}\,,
\label{eq:cbound1tau3}\\
{\left| {C_9^{\mu \tau }} \right|^2} &\le& A_{\mu\nu\bar{\nu}}\,
\dfrac{{96 \sqrt 2 {\pi ^5}{\Gamma _\tau }c_{{B_s}}^2}}{{{\alpha
      ^2}G_F^3m_\tau ^5{{\left| {{V_{tb}}V_{ts}^ \star }
	\right|}^2}}}\times \nn
&& {\rm max}\{ \Delta_{\tau\to\mu\nu\bar{\nu}}\}\times
\left( {1 +  {X_{{B_s}}}} \right) \,.
\label{eq:cbound1taunu}
\eea
For scenario 1 we obtain $A_{3\mu}^{(1)}=16$ and $A_{\mu\nu\bar{\nu}}^{(1)}=4$,
while for scenario 2 we get $A_{3\mu}^{(2)}=3$ and
$A_{\mu\nu\bar{\nu}}^{(2)}=1$.

The bounds from $\tau\to\mu\nu\bar{\nu}$ only depend on the fine-tuning measure
$X_{B_s}$, while those from $\tau\to 3\mu$ also depend on the value of $C_9^{\mu
  \mu}$ (and $C_{10}^{\mu \mu }$ in scenario 2) determined from the fit to $b\to
s\mu^+\mu^-$ data. The latter bounds disappear in the limit $C_9^{\mu \mu}\to 0$, as in this case the $Z'\mu\mu$
couplings may vanish so that the $\tau\to 3\mu$ decay does not receive
contributions from $Z^\prime$ exchange.

From the upper bounds on $C_{9,10}^{\tau\mu}$, we can finally determine
the maximal branching ratios for the LFV $B$ decays with $\tau\mu$ final
states. They are shown in Fig.~\ref{Brmax} for scenario 1 with
$X_{B_s}=20$ and $X_{B_s}=100$ (in scenario 2 they are a factor of 1/2
smaller). The kink in the curves occurs at the point where the
$C_{9,10}^{\mu\mu}$-independent constraint from $\tau\to\mu\nu\bar{\nu}$
becomes stronger than the constraint from $\tau\to3\mu$. One should
  note that the bounds presented in Fig.~\ref{Brmax}, which are given for
  $m_{Z'}=1\,$TeV, have only a weak logarithmic dependence on the $Z'$
  mass.

Comparing these results to the experimental upper limits in
\eq{BllEXP}, we see that the current experimental sensitivity is still
two orders of magnitude weaker. However, LHCb will be able to achieve
significant improvements in these channels.

In the  case of $\mu e$ final states, the stringent
bound from ${\rm{Br}}[\mu\to e\gamma]$ renders LFV $B$ decays
unobservable in the $C_9^{\mu\mu}$ region favored by current $b\!\to\! s\mu^+\mu^-$ data.
For $C_9^{\mu\mu}\to 0$, ${\rm{Br}}[B\to K^{(*)} \mu^\pm e^\mp]$ can become relevant with
its maximal size being constrained to ${\cal O}(10^{-7})$ from $\mu\to e\nu\bar{\nu}$.

\section{Conclusions\label{conclusion}}

In this article we have investigated the possible size of the branching ratios
of lepton-flavour violating $B$ decays $B_s\to\tau^\pm\mu^\mp$,
$B_s\to\mu^\pm e^\mp$, $B\to K^{(*)}\tau^\pm\mu^\mp$ and $B\to
K^{(*)}e^\pm\mu^\mp$ in generic $Z^\prime$ models.  {Motivated by the
model-independent fit to $b\to s$ transitions, we have considered two scenarios,
one with vectorial (scenario 1) and another one with purely left-handed couplings (scenario 2) of the
$Z^\prime$ to leptons. 

From $\rm{Br}(B\to K\nu\bar{\nu})$ one obtains limits on the branching ratios of LFV
$B$ decays of $8(2)\times 10^{-5}$ for scenario 1(2) simply by using
gauge invariance. However, even stronger bounds can be obtained by
combining the leptonic constraints with a limit on the amount of fine
tuning in the $B_s-\overline{B}_s$ system. For a fine-tuning of
$X_{B_s}\lesssim 100$,} we have found that still sizeable branching
ratios of ${\cal O}(10^{-6})$ are possible in both scenarios for
$\tau\mu$ final states, while for $\mu e$ final states they can only
reach ${\cal O}(10^{-7})$ in a region of parameter space disfavoured
by the current data on $B\to K^*\mu^+\mu^-$, $B_s\to\phi\mu^+\mu^-$
and $R_K$.

{\bf Note added:} During the publication process of this article new LHCb results on $B_s\to\phi\mu^+\mu^-$ were released, increasing the discrepancy compared to the SM to $3.5\,\sigma$~\cite{Aaij:2015esa}. 

\acknowledgments{A.C. thanks Julian Heeck for useful discussions. L.H. likes to thank Mitesh Patel and Marcin Chrzaszcz for 
useful discussions concerning $B\to K^{(*)}\nu\bar{\nu}$ and $\tau\to\mu \nu \bar{\nu}$, respectively. We
  are grateful to David Straub and Javier Virto for additional
  information concerning the model independent fits to $b\to s$
  transitions. We are grateful to Diego Guadagnoli for pointing out a missing factor 1/2 in the formula for $B_s\to \ell\ell^\prime$. J. R. and S.P. are supported in part by the Polish
  National Science Center under the research grant
  DEC-2012/05/B/ST2/02597. A.~C. is supported by a Marie Curie
  Intra-European Fellowship of the European Community's 7th Framework
  Programme under contract number PIEF-GA-2012-326948. U.N. is
  supported by BMBF under grant no.~05H12VKF. L.H. is supported by
  FPA2011-25948 and the grant 2014 SGR1450, and in part by
  SEV-2012-0234. J.M acknowledges also FPA2014-61478-EXP.}

%\clearpage
\bibliography{BIB}

\end{document}